# Evaluating amyloid-beta as a surrogate endpoint in trials of anti-amyloid drugs in Alzheimer's disease: a Bayesian meta-analysis

Sa Ren[1], Janharpreet Singh[2], Sandro Gsteiger[3], Christopher Cogley[2], Ben Reed[2], Keith R Abrams[4], Dalia Dawoud[5], Rhiannon K Owen[6], Paul Tappenden[1], Terrence J Quinn[7], Sylwia Bujkiewicz[2]

[1]School of Medicine and Population Health, University of Sheffield, Sheffield, UK, S1 4DA
[2]Biostatistics Research Group, Department of Population Health Sciences, University of Leicester, Leicester, UK, LE1 7RH
[3]F. Hoffman-La Roche Ltd, Konzern-Hauptsitz Grenzacherstrasse 124, 4070 Basel, Switzerland
[4]Department of Statistics, University of Warwick, Coventry, UK, CV4 7AL
[5]National Institute for Health and Care Excellence, UK, E20 1JQ
[6]Population Data Science, Swansea University Medical School, Swansea University, Swansea, UK, SA2 8PP
[7]School of Cardiovascular and Metabolic Health, University of Glasgow, Glasgow, UK, G12 8TA

## Abstract

**Background and Objectives:** The amyloid cascade hypothesis posits that amyloid-beta (Aβ) plays a central role in the pathogenesis of Alzheimer's disease (AD) and has consequently guided therapeutic strategies over the past several decades. The use of Aβ clearance to support regulatory approvals of drugs in AD remains controversial. We evaluate Aβ as a potential trial-level surrogate endpoint for clinical function in AD using a meta-analysis.

**Methods:** Data on the effectiveness of anti-amyloid monoclonal antibodies (MABs) on Aβ and clinical outcomes were identified from randomised controlled trials (RCTs) through a literature review. A Bayesian bivariate meta-analysis was used to evaluate surrogate relationships between the treatment effects on Aβ and on clinical function, with the intercept, slope and variance quantifying the trial-level association. The analysis was performed using RCT data both collectively across all MABs and separately for each anti-amyloid MAB through subgroup analysis. The latter analysis was extended by applying Bayesian hierarchical models to borrow information across treatments.

**Results:** We identified 23 RCTs with 39 treatment contrasts for seven MABs. The association between treatment effects on Aβ and Clinical Dementia Rating - Sum of Boxes (CDR-SOB) across all MABs was strong: with intercept of -0.03 (95% credible intervals: -0.16, 0.11), slope of 1.41 (0.60, 2.21) and variance of 0.02 (0.00, 0.05). For individual treatments, the surrogate relationships were suboptimal, displaying large uncertainty. The use of hierarchical models considerably reduced the uncertainty around key parameters, narrowing the intervals for the slopes by an average of 71% (range: 51%–95%) and for the variances by 28% (7%–65%).

**Discussion:** Our results suggest that Aβ is a potential surrogate endpoint for CDR-SOB when assuming a common surrogate relationship across all MABs. However, the surrogate relationships for individual treatments were uncertain. When allowing for information-sharing across treatments, the surrogate relationships improved, but only for lecanemab and aducanumab was the improvement sufficient to support a surrogate relationship. This suggests that the strength of the surrogate relationship between the effects on Aβ and CDR-SOB varies across anti-amyloid MABs and may not necessarily hold for new therapies. Data from a larger number of trials would be required to fully assess this.

**Keywords:** Alzheimer's disease; amyloid-beta; clinical outcomes; surrogate endpoint; meta-analysis

## 1. Introduction

Considerable research has focused on the development of monoclonal antibodies (MABs) aiming to inhibit the production of, or activate the clearance of, amyloid-beta (Aβ) in patients with Alzheimer's disease (AD). This research has culminated in the completion of a number of randomised controlled trials (RCTs) assessing anti-Aβ drugs, including: aducanumab[1]; lecanemab[2]; donanemab [3]; and gantenerumab[4], which vary in their mechanism of action. In particular, aducanumab and lecanemab select for soluble aggregated forms of Aβ, whilst other MABs target Aβ monomers (e.g., solanezumab), or do not discriminate between Aβ forms (e.g., bapineuzumab and crenezumab) [5,6].

The US Food and Drug Administration (FDA) granted licensing approval for aducanumab, based on trial evidence demonstrating treatment efficacy in terms of reducing Aβ plaques in the brain, which was considered a surrogate endpoint for clinical benefit [7,8]. This generated controversy and criticism of FDA licensing approvals due to a lack of evidence on the surrogate relationship between the treatment effect on Aβ and clinical benefit measured on cognitive function [9–11]. Despite this controversy, another FDA accelerated approval was granted for lecanemab, based on evidence of Aβ reduction [12].

More recently, European Medicines Agency (EMA) granted marketing authorization for lecanemab, only after a re-examination process following earlier refusal due to limited effect on the cognitive function and concerns over side effects including amyloid-related imaging abnormalities (ARIA). National Institute for Health and Care Excellence (NICE) declined to recommend lecanemab for reimbursement in England as it was not cost-effective. Although a considerable reduction in Aβ has been observed for MABs, there is lack of evidence of their long-term clinical benefit.

At the time of the FDA approval of aducanumab, only limited evidence existed about the association between Aβ and cognitive function, based on a relatively small cohort study [13]. However, no evidence was available about the association between the treatment effects on these two outcomes. Ackley *et al.* [14] performed an instrumental variable analysis based on data from a meta-analysis of 14 trials in AD, which found that a reduction in Aβ was not associated with a positive effect on the Mini Mental State Examination (MMSE). Later, Pang *et al.* [15] updated their meta-analysis and concluded that amyloid reduction was associated with three commonly used clinical outcomes; MMSE, Clinical Dementia Rating - Sum of Boxes (CDR-SOB), and Alzheimer's Disease Assessment Scale - Cognitive Subscale (ADAS-Cog). The instrumental variable method adopted in the above two papers was based on the assumption that the cognitive improvement is mediated by amyloid reduction. Recently, Ackley *et al.* [16] updated their previous analysis using new data from additional four RCTs and found a small effect of amyloid reductions on CDR-SOB. The updated analysis was performed using Bayesian approach but was limited as all prior distributions, representing different beliefs, had the same variance informed by the previous meta-analysis. Wang *et al.* [17] carried out a meta-analysis of data from six trials assessing four different MABs, which found a positive correlation between a decrease in Aβ measured at 52 weeks and a decrease in CDR-SOB at 78 weeks. In another recent study, Wang *et al.* [18] carried out a latent class analysis using patient-level data, which suggested that Aβ reduction was associated with improved clinical outcomes. However, this study was limited to data from a single trial of only two MABs (solanezumab and gantenerumab) and in terms of the sample size and lack of randomisation to the classes determined by the model (amyloid-no-change, amyloid reduction and amyloid growth).

In this paper, we carry out a Bayesian meta-analysis of RCTs of anti-Aβ MABs to evaluate Aβ as a trial-level surrogate endpoint for clinical outcome. We evaluate surrogacy patterns between the treatment effects on the two outcomes across all MABs trials, and within subgroups of RCTs of individual MABs to acknowledge the potential impact of their different mechanisms of action on the surrogate relationships. We utilise a Bayesian framework to allow for borrowing of information on the surrogate relationship across treatments.

## 2. Methods

### *2.1 Data sources, extraction and outcomes*

We performed a literature search to identify recent systematic reviews of RCTs assessing the effectiveness of anti-Aβ monoclonal antibodies (MABs) in patients with AD, followed by a search of the clinical trials database ClinicalTrials.gov. Our evidence base consisted of a comprehensive set of relevant RCTs, conducted in any phase, that reported treatment effects on both Aβ and clinical outcome, as included in the identified systematic reviews.

Data extraction from each trial was undertaken by two authors (SR and JS) and checked independently by two authors (CC and BR). Data were extracted from the corresponding trial publication (and its associated supplementary materials), or from ClinicalTrials.gov when they were not reported in a trial publication. In cases where the treatment effect (and the associated standard error) was not reported numerically, these data were digitised from reported graphs using digitiser tools.

Data were extracted on the treatment effects on Aβ and clinical function. Data for the effect on Aβ were extracted on positron emission tomography (PET) standardised uptake value ratio (SUVR) as well as the Centiloid scale. Unreported data on one of the two scales for Aβ were imputed utilising mapping equations identified from the literature [19–22], which are described in Appendix A. Treatment effects on clinical function were obtained for the following outcomes: CDR-SOB, MMSE, and ADAS-Cog. The treatment effect on an outcome for each treatment contrast (MAB versus placebo) was defined as the difference in the change from baseline of the outcome measure between the active and control treatment arms. Data were extracted for the latest time point at which results were reported on both outcomes, and also at an earlier time point for Aβ and a later time point for clinical function if available. Additional data were extracted on the following factors: amyloid-related imaging abnormalities (ARIA), the most common adverse event observed in Phase III trials of aducanumab, [23] and the proportion of carriers of at least one ε4 allele of apolipoprotein E (APOE) biomarker in each treatment arm.

## 2.2 Statistical analyses

We adopted a Bayesian bivariate meta-analysis model (allowing for the incorporation of multi-arm trials) to perform trial-level evaluation of Aβ as a surrogate endpoint for clinical function, following the method of Daniels & Hughes [24]. The surrogate relationship between the treatment effects on Aβ level and on the clinical function was described in the form of a regression equation. The strength of the association was evaluated based on the criteria set out by Daniels & Hughes, who consider a perfect surrogate relationship when: (i) the intercept is zero, meaning that a null effect on the surrogate endpoint should imply a null effect on the final clinical outcome; (ii) slope is non-zero, which signified evidence of an association between effects on the surrogate endpoint and the final outcome, and (iii) the variance (of the treatment effect on the clinical outcome conditional on the effect on the surrogate endpoint) is zero, implying that the treatment effect on the final outcome could be predicted perfectly from the effect on the surrogate endpoint.

As a base case analysis, we assessed the surrogate relationship across all trials in the data set, irrespective of MAB treatment. Additionally, we assessed the trial-level surrogate relationship for each treatment separately, by carrying out subgroup analyses. We then extended the analyses for individual treatments utilising Bayesian hierarchical models [25]. The method allows for borrowing information about the surrogate relationship across treatments whilst retaining the ability to differentiate between the strong and weak surrogate association patterns for the individual treatments. Both full exchangeability and partial exchangeability hierarchical models have been adopted for different strength of borrowing [25]. We performed a range of sensitivity analyses by selecting different measures of treatment effect on clinical function and Aβ, including outcomes reported at different follow up time, considering the impact of the choice of prior distribution on the conditional variance, and adjusting the surrogate relationship for the treatment effect on ARIA and for the proportion of APOE carriers. We performed leave-one-out cross validation to assess the predictive value of Aβ effect on the effect of clinical function [26].

Within-study correlation between Aβ and clinical function is needed for each study to populate the bivariate meta-analytic model; however, this was only reported for two RCTs [27] at the arm level. For the studies that did not report the correlation, the average of the two reported correlations was used, assuming the same correlation across the remaining trials [24].

We implemented the models using the WinBUGS 1.4.3 software package to estimate model parameters using Markov chain Monte Carlo (MCMC) simulation [28]. A total 100,000 iterations with a burn-in of 50,000 and thinning of 10 were used to estimate the model parameters. Convergence was checked by visually assessing the trace plots and autocorrelation. The results are presented as means with 95% credible intervals (CrIs). Additional analyses, including data management and graphics, were conducted using R software [29].

## 3. Results

### 3.1 Literature review

We identified 13 meta-analyses seeking to synthesise evidence from RCTs on the efficacy of anti-Aβ drugs for AD [30–42]. These meta-analyses focused on pooling effectiveness data on each outcome individually. Three of them synthesised the effects of anti-Aβ drugs on both PET Aβ and clinical outcome measures (ADAS-Cog, CDR-SOB and MMSE) [30–32]. Although the results of these meta-analyses showed an overall significant effect in reducing Aβ, the effect on the clinical outcome measures varied across the meta-analyses. There were six meta-analyses [33–38] synthesising the treatment effects on clinical outcomes but not PET Aβ. One review reported treatment effects on PET Aβ but not on clinical outcomes [39]. Additionally, there were three systematic reviews comparing the efficacy of different MABs on various outcome measures using network meta-analysis [40–42]. Some further details pertaining to the conclusions of these meta-analyses can be found in Appendix B. Following detailed review of the identified meta-analyses, we found the systematic review by Jeremic *et al.* [38] most up to date and complete, as it comprised all relevant trials included in all the other systematic literature reviews. Thus, the trials identified by Jeremic *et al.* [38], formed the evidence base for our research. Our literature search also identified a review of ongoing Phase II/III trials assessing the efficacy of anti-Aβ drugs in AD [43]. A review of these trials along with a search of ClinicalTrials.gov data base, which were conducted to ensure our evidence base was up to date, did not identify any additional completed trials.

3.1.1 Data set

Figure 1 presents the data set used to perform the meta-analysis evaluating the surrogate relationship. These data represent the observed effects on Aβ PET and CDR-SOB, comparing the active treatment arms with the placebo control arm in each trial. The effects are grouped by treatment. The data were obtained from 23 trials and 39 treatment contrasts reporting effects on both Aβ PET and CDR-SOB.

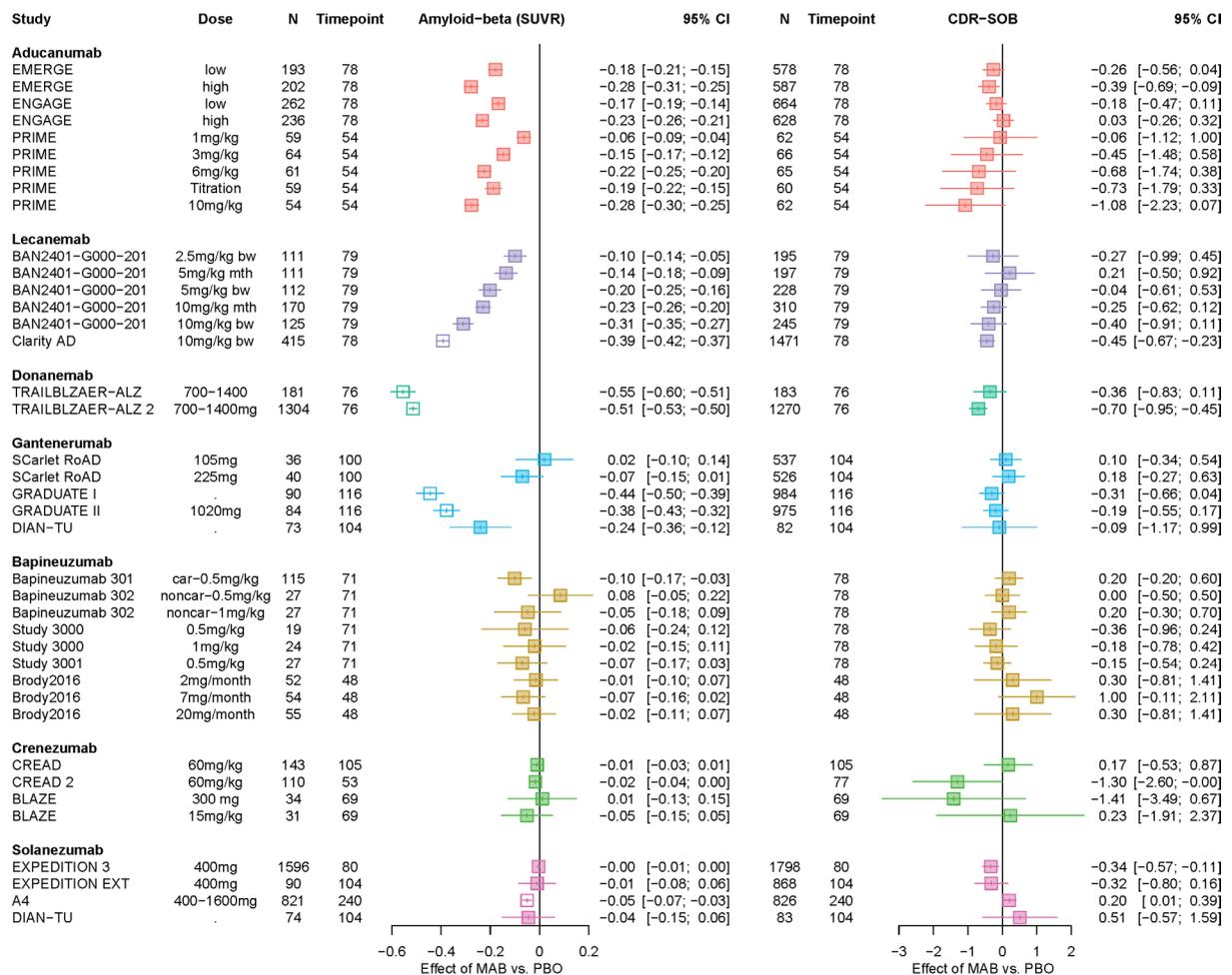

Figure 1. Forest plot illustrating the treatment effects of anti-amyloid-beta monoclonal antibodies (MABs) on amyloid-beta levels (measured via positron emission tomography (PET) scan, on the standardised uptake value ratio (SUVR) scale) and the Clinical Dementia Rating – Sum of Boxes (CDR-SOB) outcome. The treatment effects represent the difference in change from baseline between MAB and placebo (PBO). Estimates shown as empty squares (no filled colour) were imputed by applying a conversion formula based on the radioactive tracer used in the PET scan, for trials where the effect on amyloid-beta was reported on the Centiloid scale alone.

For aducanumab, lecanemab and donanemab, trials demonstrated statistically significant effects on Aβ PET across different doses. There was evidence of a dose-response relationship, with a higher dose corresponding to a larger effect. However, for most of these treatments' doses there was no evidence of a statistically significant effect on CDR-SOB (exceptions were the high-dose arms in the EMERGE trial, Clarity AD trial and TRIALBLAZER-ALZ 2). Despite this, the point estimates indicate that a larger effect on Aβ PET is associated with a larger effect on CDR-SOB. There was larger evidential uncertainty for the other treatments (gantenerumab, bapineuzumab, crenezumab, solanezumab), where a statistically significant effect was apparent for five (out of 22) treatment contrasts on Aβ PET and for two (out of 22) treatment contrasts on CDR-SOB. Consequently, it is unclear whether there is an association between the effects on Aβ PET and CDR-SOB for these treatments.

*3.2 Surrogate endpoint evaluation*

3.2.1 Surrogate relationships across all trials of MABs

Data from 23 identified studies with 39 treatment contrasts reporting the treatment effects on both Aβ and CDR-SOB were included. They comprised 14 two-arm studies, six three-arm studies, one four-arm study and two six-arm studies.

Figure 2 shows a bubble plot representing reported treatment effects on Aβ and CDR-SOB from all included studies with bubble size corresponding to the number of patients with reported CDR-SOB. The regression line represents the surrogate relationship between treatment effects on Aβ and CDR-SOB from the analysis of data across all MABs trials. The relationship across all MABs was strong, characterised by an intercept close to zero at -0.03 (95% CrI: -0.16, 0.11), a positive slope of 1.41 (95% CrI: 0.60, 2.21) and a small conditional variance of 0.02 (95% CrI: 0.00, 0.05).

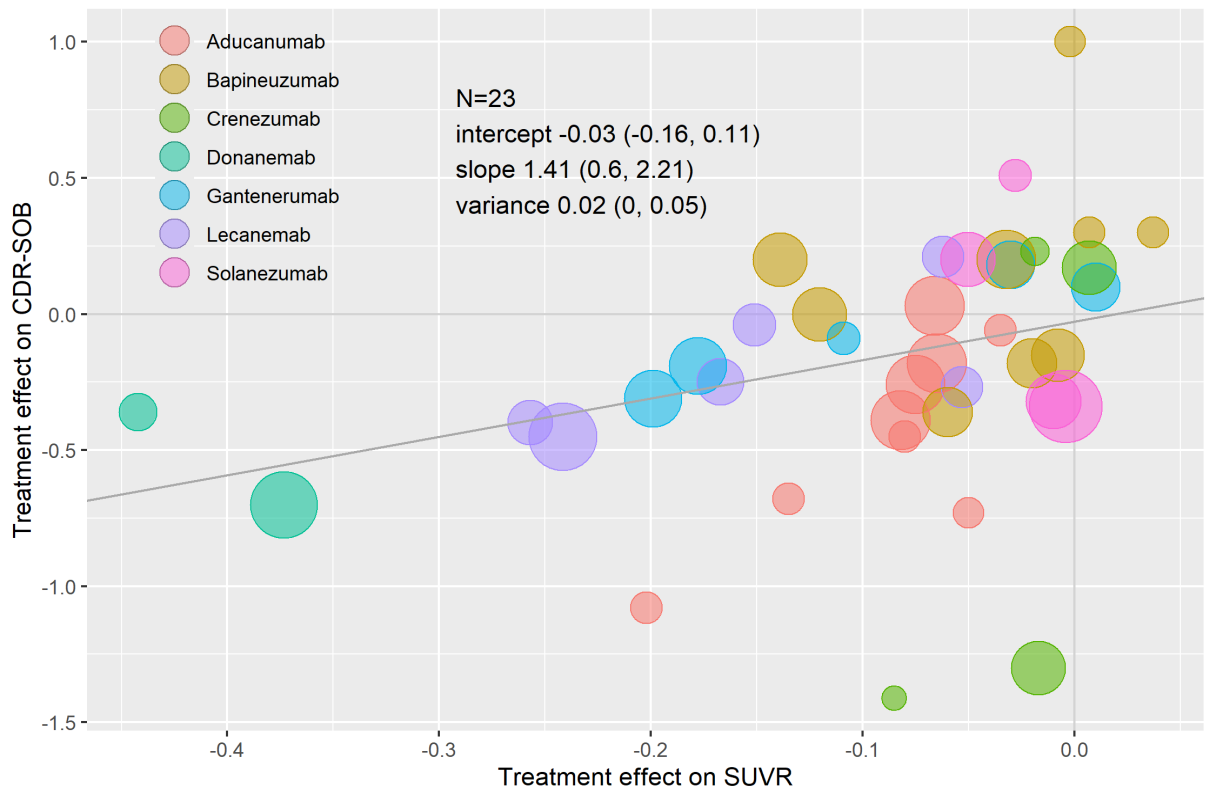

Figure 2. Bubble plot of the surrogate relationship between treatment effects on amyloid-beta measured on the standardised uptake value ratio (SUVR) scale and Clinical Dementia Rating – Sum of Boxes (CDR-SOB) with treatment effects on amyloid-beta SUVR reported at earlier time points. The mean estimates (95% credible intervals) of intercept, slope and conditional variance were obtained from Daniels and Hughes model. The bubble size corresponds to the number of patients with CDR-SOB reported.

Leave-one-out cross-validation was performed to evaluate the predictive value of Aβ as a surrogate endpoint for CDR-SOB. The cross-validation across all trials showed a good coverage rate with 95% of the predicted intervals including the observed estimates of the effects on CDR-SOB. A forest plot showing the observed effects and the predicted effects of CDR-SOB for each study can be found in Figure S1 of Appendix C1.

Sensitivity analyses were conducted to further explore the surrogate relationships across all MABs, utilising different measures of the treatment effect on Aβ and the clinical function. Two clinical outcomes were considered: ADAS-Cog and MMSE. When using ADAS-Cog as the measure of clinical function, data from 20 trials and 31 treatment contrasts were available. The surrogate relationship between treatment effects on Aβ SUVR and ADAS-Cog was estimated to be weak with much larger conditional variance, 0.06 (95% CrI: 0.00, 0.23), compared to the analysis using CDR-SOB as the final clinical outcome. The analysis of data on the clinical function measured by MMSE included 13 trials and 22 contrasts. We found lack of evidence of surrogate relationship between treatment effects on Aβ SUVR and MMSE, with the 95% credible interval for the slope including both positive and negative values. Both sets of results are presented in Figures S2 and S3 of Appendix C2.

Another set of analyses considered outcomes reported at different follow up times. The main analysis, reported above, included data with the follow up at the earliest reported treatment effect on Aβ. When using the treatment effects on Aβ reported at the same time points as the clinical outcome, the surrogate relationship between treatment effects on Aβ SUVR and CDR-SOB was also strong, as presented in Figure S4 of the Appendix C2. The surrogate relationship between treatment effects on Aβ using the Centiloid scale (as an alternative measurement unit to SUVR) and CDR-SOB was estimated to be strong. This result was based on data from all 23 trials and 39 contrasts. Further details of the analysis can be found in Figure S5 of the Appendix C2.

3.2.2 Surrogate relationships for individual treatments

The surrogate relationships across trials within individual treatments were also explored. The results using subgroup analysis, and a full exchangeability hierarchical model allowing for information sharing across the treatments, are presented in Figure 3. Each column shows the estimated slope (red), intercept (green), and conditional variance (blue) for each treatment in turn**.** The surrogate relationships estimated from the subgroup analyses were weak, with wide CrIs around the association parameters for all individual treatments. The use of the full exchangeability hierarchical model, through borrowing of information from other treatments, allowed us to estimate the key parameters with much higher precision. The reduction in the width of 95% CrI was on average 71% (ranging between 51%-95%) for the slope and 28% (ranging between 7%-65%) for the conditional variance, when comparing the results from the hierarchical model with subgroup analyses respectively.

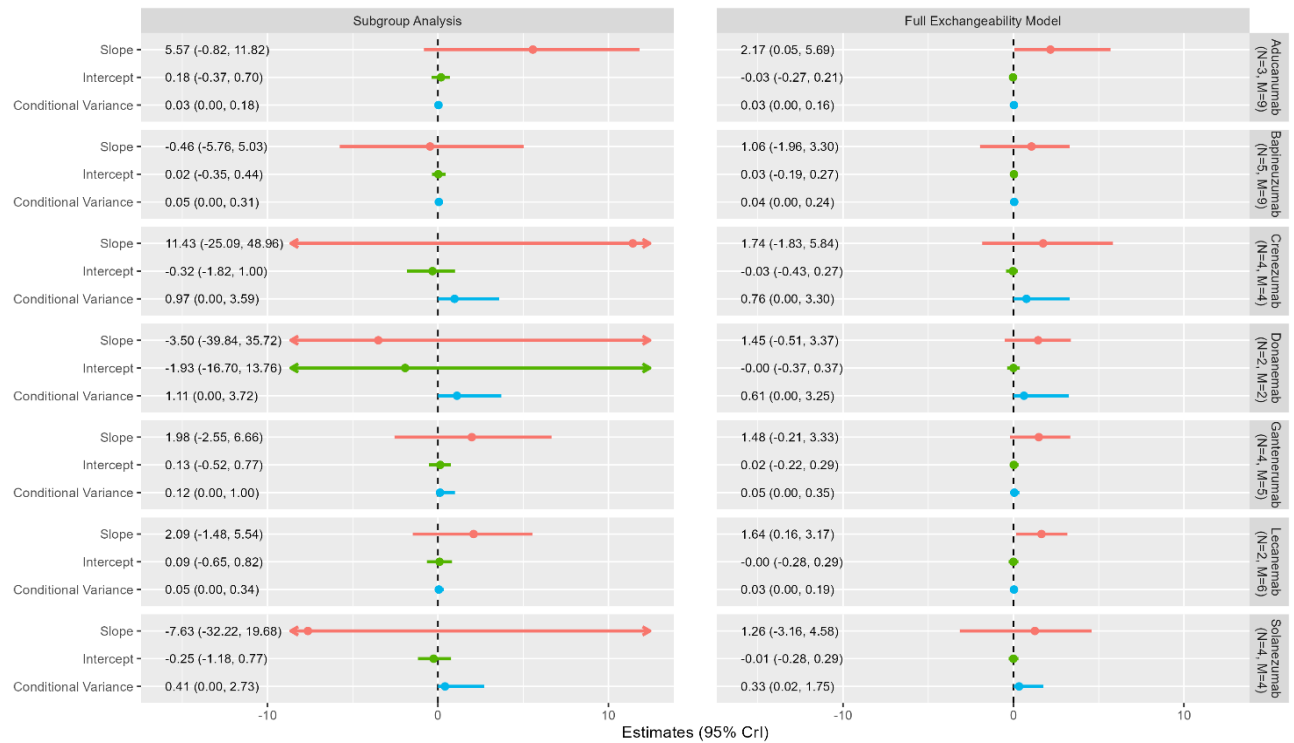

Figure 3. Forest plot of estimates of slope (red), intercept (green) and conditional variance (blue) for amyloid-beta (on the standardised uptake value ratio (SUVR) scale) as a surrogate for Clinical Dementia Ratio – Sum of Boxes (CDR-SOB) using treatment effects on amyloid-beta SUVR at earliest reported time points. Each column represents a different model, and each row corresponds to a different treatment. N represents the number of studies included and M represent the total number of contrasts included.

For example, the surrogate relationship across treatment effects on SUVR and CDR-SOB from trials of lecanemab alone was weak, with a slope of 2.09 (95% CrI: -1.48, 5.54) and a conditional variance of 0.05 (95% CrI: 0.00, 0.34). This surrogate relationship was stronger when sharing information from trials of other MABs, with the hierarchical model resulting in a slope of 1.64 (95% CrI: 0.16, 3.17) and a conditional variance of 0.03 (95% CrI: 0.00, 0.19), corresponding to the reduction in the width of CrI by 57% for slope and 46% for conditional variance, respectively. Similar results were obtained for aducanumab. The slope was estimated to be 5.57 (95% CrI: -0.82, 11.82) and the conditional variance of 0.03 (95% CrI: 0, 0.18) when only data from aducanumab trials were used, while a positive slope of 2.17 (95% CrI: 0.05, 5.69) and a smaller conditional variance of 0.03 (95% CrI: 0.00, 0.16) were obtained when information from other MABs was borrowed using the hierarchical model. The reduction in the width of CrI was 55% for slope and 7% for conditional variance, respectively. These results suggest a moderate surrogate relationship between the treatment effect on Aβ measured by SUVR and CDR-SOB for lecanemab and aducanumab, albeit still with a high level of uncertainty around these key parameters.

There was a lack of evidence of a surrogate relationship between the effects on Aβ SUVR and CDR-SOB for all the other drugs, with the 95% CrIs for all estimated slopes including zero and large conditional variances, in particular for crenezumab, donanemab and solanezumab. The use of the hierarchical model led to the reduction in the width of CrIs, ranging from 51% to 95% for the slope, and from 8% to 65% for the conditional variance. However, despite the increased precision gained from borrowing of information, the surrogate relationships were still uncertain for these individual treatments.

Similar results across all treatments were obtained from the partial exchangeability model, but with a slightly higher level of uncertainty around the estimates for the slope, intercept and conditional variance. These results are presented in Figure S7 of Appendix C3.

Leave-one-out cross-validation was also performed to evaluate the predictive value of Aβ using the hierarchical models. The full exchangeability model demonstrated 100% coverage, which is likely associated with inflated predicted intervals due to increased uncertainty in the key surrogacy parameters. A forest plot showing the observed effects and predicted effects of CDR-SOB for each study can be found in Figure S8 of Appendix C4.

## 4. Discussion

We evaluated Aβ level as a putative surrogate endpoint for clinical function using Bayesian meta-analysis models for surrogate endpoint evaluation. The evidence included in the meta-analysis was extracted from RCTs that reported both the treatment effects on Aβ level and on clinical function. 23 RCTs with 39 treatment contrasts for seven MABs were identified and analysed collectively. Results from the meta-analysis of 23 RCTs showed that an effect on Aβ was a potential surrogate endpoint for the effect on CDR-SOB when assuming a common surrogate relationship for all included treatments. There was a lack of evidence of a surrogate relationship between treatment effects on Aβ and MMSE based on the meta-analysis of 13 trials. The results from the meta-analysis of 20 RCTs suggested that the surrogate relationship between treatment effects on Aβ and ADAS-Cog was relatively weak.

Recently, results from Pang *et al.* [15], Ackley *et al.* [16] and Wang *et al.* [17] also suggested a significant association between the Aβ reduction and CDR-SOB improvement, consistent with our results. Our meta-analysis combined all available information related to treatment effects on Aβ and clinical function from 23 RCTs using a Bayesian surrogate evaluation model, providing a more robust estimate of the association between the effects on Aβ and the effects on clinical function. Two earlier meta-analyses [14,15] found no association between Aβ reduction and MMSE improvement based on published antibody data, which aligns with our results. Evidence related to treatment effects on MMSE was limited; only 13 of the 23 identified trials reported treatment effect on MMSE, most of which did not achieve a big reduction in Aβ. Also, MMSE has been found to have low sensitivity to cognitive deterioration [44]. These factors may have contributed to the parameters of surrogate relationship obtained with high uncertainty.

The surrogate relationship between treatment effects on Aβ level and change in ADAS-Cog was found to be suboptimal with statistically meaningful slope but moderate conditional variance. Among the 20 included RCTs, four different variants of questionnaires were used to measure ADAS-Cog, which may have introduced additional heterogeneity and, therefore, additional uncertainty of the results. Results from Pang *et al.* [15] also showed a significant effect of reduction on ADAS-Cog with a wide confidence interval.

For each treatment, we also conducted subgroup analysis using only data extracted from that specific treatment. The surrogate relationships between treatment effects on Aβ and CDR-SOB for each individual treatment were estimated to be highly uncertain due to a small number of trials. With the use of Bayesian hierarchical model, individual surrogacy parameters were estimated with much improved precision by borrowing information from trials on other treatments. Statistically meaningful slopes were estimated for aducanumab and lecanemab when allowing for borrowing of information, suggesting a potential surrogate relationship for the two treatments. There was a lack of meaningful improvement in surrogate relationship for other treatments. This may be partly due to small number of trials for each treatment, but the results were also suboptimal for Bapineuzumab where the amount of data was comparable if not exceeding the quantity for lecanemab and aducanumab. This suggests that the strength of the surrogate relationship between the effects on Aβ and CDR-SOB varies across anti-amyloid MABs and may not necessarily hold for new therapies. This also highlights the robustness of the hierarchical models.

One limitation of our study was existence of missing data on the treatment effects on Aβ measured on SUVR scale. We used mapping equations to impute the Aβ measurements for six studies that did not report the effect on the SUVR scale and reported Aβ on the Centiloid scale instead. While the mapping equations allow us to incorporate all the relevant evidence in the analysis, this approach may have also introduced uncertainty into the results. To address this concern, a sensitivity analysis excluding imputed estimates was carried out and the results are provided in Figure S6 of Appendix C2. Additionally, there was heterogeneity in the follow-up time across studies, as mentioned earlier. As discussed in Section 3.2.1, a sensitivity analysis to the length of the follow up time was also carried out, which showed similar results as the main analysis. Furthermore, the use of different tracers measuring PET Aβ levels across studies may have contributed to additional between-study heterogeneity of the effects on Aβ and, as a result, increased uncertainty in estimating key surrogacy parameters.

Future work should focus on combing data from multiple sources such as cohort studies and potentially single arm trials to maximize the use of all available information and enhance precision in evaluating surrogate endpoints in AD. Information from these studies can also help inform key model parameters [45]. However, such analysis would require a careful consideration of potential bias and confounding. Additionally, previous research has found associations between Aβ reduction, ARIA rate, APOE-ε4 and clinical efficacy [17], suggesting that the proportion of APOE carriers and the presence of ARIA may need to be considered when assessing surrogate endpoints. Although our initial sensitivity analyses have shown that these factors minimally affect surrogate relationships, further research is needed to investigate their potential influence in greater depth and refine the evaluation process. The benefit of the Bayesian hierarchical models may have been limited for treatments with a small number of trials, for example, donanemab. Data from a larger number of trials would be required to fully assess the surrogate relationships across other anti-amyloid MABs.

Our findings suggest that Aβ reduction could potentially serve as a surrogate endpoint in clinical trials, thereby accelerating the evaluation of novel AD therapies. By using Aβ reduction as an early indicator of clinical efficacy, researchers may be able to speed up the drug development processes and reduce the time required to bring new therapies to the market. However, it is important to recognize that while Aβ reduction is associated with clinical improvement overall, it does not guarantee such an improvement; particularly in individual treatments including therapies developed in the future. Relying solely on Aβ as a surrogate endpoint may overlook other relevant pathophysiological processes contributing to AD. Aβ reduction should be viewed as part of a broader strategy to understand AD.

## Acknowledgements

This research was funded by the Medical Research Council [MR/T025166/1]. SB and JS were also supported by Leicester NIHR Biomedical Research Centre (BRC). The views expressed are those of the author(s) and not necessarily those of the NIHR or the Department of Health and Social Care. The authors are grateful to the authors of the review by Jeremic et al for providing additional information about the studies from their review.

19. Navitsky, M. *et al.* Standardization of amyloid quantitation with florbetapir standardized uptake value ratios to the Centiloid scale. *Alzheimer's & Dementia* 14, 1565–1571 (2018).

20. Hanseeuw, B. J. *et al.* Defining a Centiloid scale threshold predicting long-term progression to dementia in patients attending the memory clinic: an [18F] flutemetamol amyloid PET study. *Eur J Nucl Med Mol Imaging* 48, 302–310 (2021).

21. Rowe, C. C. *et al.* 18F-Florbetaben PET beta-amyloid binding expressed in Centiloids. *Eur J Nucl Med Mol Imaging* 44, 2053–2059 (2017).

22. Klunk, W. E. *et al.* The Centiloid Project: Standardizing quantitative amyloid plaque estimation by PET. *Alzheimer's & Dementia* 11, 1-15.e4 (2015).

23. Salloway, S. *et al.* Amyloid-Related Imaging Abnormalities in 2 Phase 3 Studies Evaluating Aducanumab in Patients With Early Alzheimer Disease. *JAMA Neurology* 79, 13–21 (2022).

24. Daniels, M. J. & Hughes, M. D. Meta-analysis for the evaluation of potential surrogate markers. *Stat Med* 16, 1965–1982 (1997).

25. Papanikos, T. *et al.* Bayesian hierarchical meta-analytic methods for modeling surrogate relationships that vary across treatment classes using aggregate data. *Statistics in Medicine* 39, 1103–1124 (2020).

26. Bujkiewicz, S., Achana, F., Papanikos, T., Riley, R. D. & Abrams, K. R. Multivariate meta-analysis of summary data for combining treatment effects on correlated outcomes and evaluating surrogate endpoints. *NICE DSU Technical Support Document 20*.

27. Budd Haeberlein, S. *et al.* Two Randomized Phase 3 Studies of Aducanumab in Early Alzheimer's Disease. *J Prev Alzheimers Dis* 9, 197–210 (2022).

28. Spiegelhalter, D; Thomas, A; Best, N; Lunn, D. *WinBUGS User Manual Version 1.4*. (MRC Biostatistics Unit, 2003).

29. R Core Team. R: A language and environment for statistical computing.

30. Avgerinos, K. I., Ferrucci, L. & Kapogiannis, D. Effects of monoclonal antibodies against amyloid-β on clinical and biomarker outcomes and adverse event risks: A systematic review and meta-analysis of phase III RCTs in Alzheimer's disease. *Ageing Res Rev* 68, 101339 (2021).
20

## Appendix A. Conversion of data on Aβ PET between Centiloid and SUVR scales

In trials assessing anti-Amyloid-β (Aβ) drugs, the treatment effect is commonly assessed in terms of the reduction in Aβ vlolume within the brain measured via positron emission tomography (PET) imaging. The PET images are inspected to quantify the Aβ volume as a standardized uptake value ratio (SUVR). PET imaging involves the injection of a radioactive tracer which may differ from trial to trial, hindering the comparison of SUVR values across trials. To facilitate cross-trial comparisons, methods have been proposed to convert SUVR values to a standard Centiloid scale [1].

In our evidence base, the majority of trials reported data on the SUVR scale, but there were a few trials which reported data on the Centiloid scale only. To make the most efficient use of this evidence, we converted these data from the Centiloid scale to the SUVR scale. A number of studies have implemented the conversion from the SUVR scale to the Centiloid scale based on the tracer used for Aβ PET imaging; florbetapir [2], florbetaben [3], flutemetamol [4]. We applied the regression equations reported in these studies to make the conversion of data from the Centiloid scale to the SUVR scale, according to the tracer used in each trial. Specifically, we used the following equations:

- Florbetapir:    Centiloid = 183 x SUVR – 177

- Florbetaben:    Centiloid = 153.4 x SUVR -154.9

- Flutemetamol:    Centiloid = 116 x SUVR – 113.9

These equations were originally applied to convert SUVR values recorded at a particular point in time. In our evidence base, the data are reported as the change from baseline so we apply each equation without the intercept terms (since these terms would cancel out when applying the equation to data at each time point, i.e., at baseline and follow-up). For trials in which the multiple tracers were used (e.g., florbetapir and florbetaben), we make the conversion using the average of the regression equations (i.e., averaging the coefficients).

## Appendix B. Literature review

Three studies include the effect of anti-Aβ drugs on both PET Aβ and clinical outcome measures (ADAS-Cog, CDR-SOB and MMSE) [5–7]. Although the results of these meta-analyses showed an overall significant effect in reducing Aβ, the effect on the clinical outcome measures varied across the meta-analyses. Avgerinos et al. [5] found a small positive effect of anti-Aβ drugs on both ADAS-Cog and MMSE, but no effect was evident on CDR-SOB based on 17 studies. Lacorte et al. [6] identified and found a negative effect of anti-Aβ drugs on CDR-SOB after performing a synthesis. Lyu et al. [7] found significant effects on ADAS-Cog, CDR-SOB and MMSE.

There were six meta-analyses [8–13] focusing on different clinical outcomes. Jeremic et al. [8] found a positive effect of anti-Aβ drugs on ADAS-Cog, MMSE, and CDR-SOB. Richard *et al.* [9] found a lack of evidence of an effect on ADAS-Cog based on six trials. Fernandez *et al.* [10] found a small but statistically significant effect on ADAS-Cog and MMSE based on 12 RCTs. Villain *et al.* [11] found a statistically significant effect on ADAS-Cog and CDR-SOB based on evidence from 4 RCTs demonstrating high level of amyloid clearance, but no effect was found on MMSE. After conducting a Bayesian random effects meta-analysis, Teipel *et al.* [12] found a moderate to small but statistically meaningful effect (with 95% credible intervals excluding zero) on CDR-SOB based on eight studies. Holdridge *et al.* [13] found a statistically significant effect across various functional outcome measures based solely on three phase III studies of solanezumab.

# Appendix C. Further Results

Appendix C1. Cross validations and predictions for the analysis of surrogate relationships across all trials of MABs

Leave-one-out cross-validation was performed to evaluate the predictive value of Aβ as a surrogate endpoint for CDR-SOB. The Daniels and Hughes model showed a good coverage with 95% of the predicted intervals including the observed estimates of the effects on CDR-SOB. The average absolute difference between the observed and predicted effect estimates was 0.32, and the average ratio of the width of intervals between the predicted and observed treatment effect was 1.24. A forest plot showing the observed effects and predicted effects of CDR-SOB for each study are presented in Figure S1.

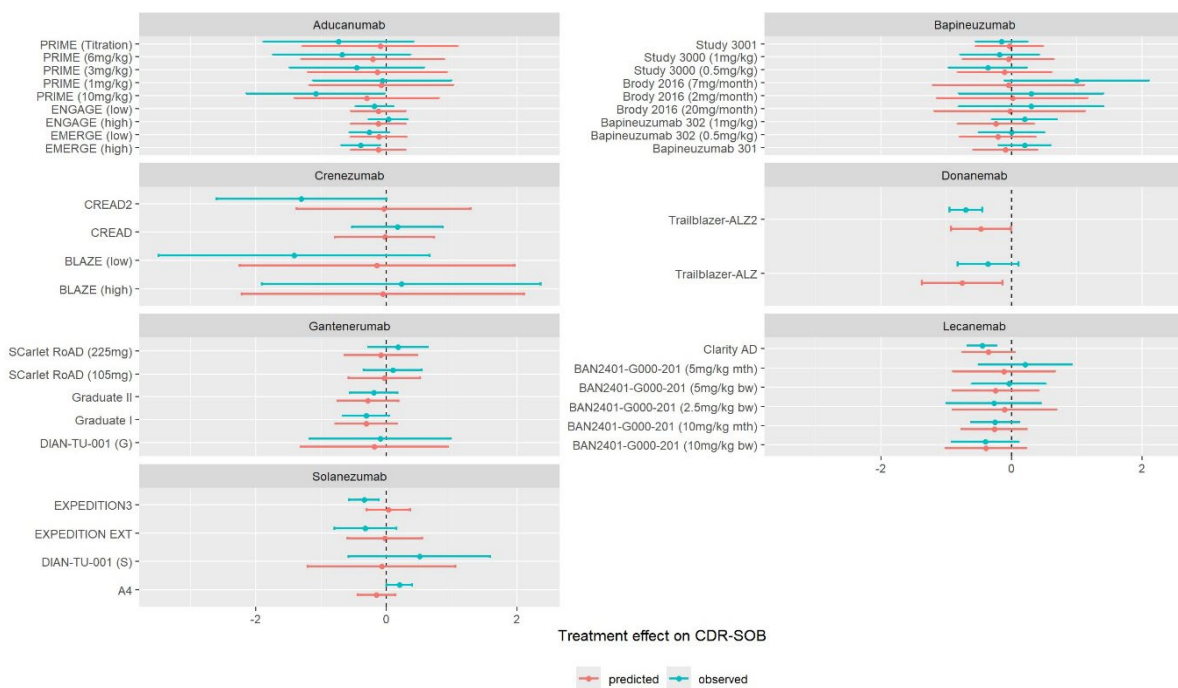

Figure S1. Forest plot illustrating the observed treatment on Clinical Dementia Rating – Sum of Boxes (CDR-SOB) and the corresponding predicted effect on CDR-SOB using Daniels and Hughes model.

Appendix C2. Further analyses for evaluating surrogate relationship across all trials of MABs

Analyses were also conducted to explore the overall surrogate relationship between treatment effects on Aβ SUVR and other clinical outcomes. The surrogate relationship between treatment effects on Aβ SUVR and ADAS-Cog was estimated to be weak, with slope 3.71 (95% CrI: 1.44, 6.05) and 0.06 (95% CrI: 0, 0.28). This analysis is based on data extracted from 20 studies (31 treatment contrasts) that have collected both treatment effects on Aβ and ADAS-Cog. Among the 20 include studies, five studies reported ADAS-Cog11, two studies reported ADAS-Cog12, eight studies reported ADAS-Cog13 and five studies reported ADAS-Cog14. 13 studies with 22 contrasts reported both treatment effects on Aβ and MMSE. The surrogate relationship between treatment effects on Aβ SUVR and MMSE was uncertain, with slope -1.14 (95% CrI: -2.84, 0.49) and conditional variance 0.06 (95% CrI: 0, 0.23). The bubble plot and regression line are presented in Figure S2 and Figure S3 respectively.

When the treatment effects were reported at the same time, the surrogate relationship between Aβ SUVR and CDR-SOB was also strong, with slope 0.93 (95% CrI: 0.4, 1.47) and conditional variance 0.02 (95% CrI: 0, 0.05). The bubble plot and regression line are presented in Figure S4. In some trials, Aβ levels were measured on a centiloid scale instead of the SUVR scale. The surrogate relationship between treatment effects on Aβ using centiloid measure and CDR-SOB was found to be strong, with slope 0.01 (95% CrI: 0, 0.01) and conditional variance 0.02 (95% CrI: 0, 0.05). The bubble plot and regression line are presented in Figure S5. Six studies were identified with both treatment effects on Aβ centiloids and CDR-SOB reported. Despite the limited number of studies included, the surrogate relationship across the six included studies was found to be moderate with slope 0.01 (95% CrI: 0, 0.02) and conditional variance 0.05 (95% CrI: 0, 0.35). The bubble plot and regression line are presented in Figure S6.

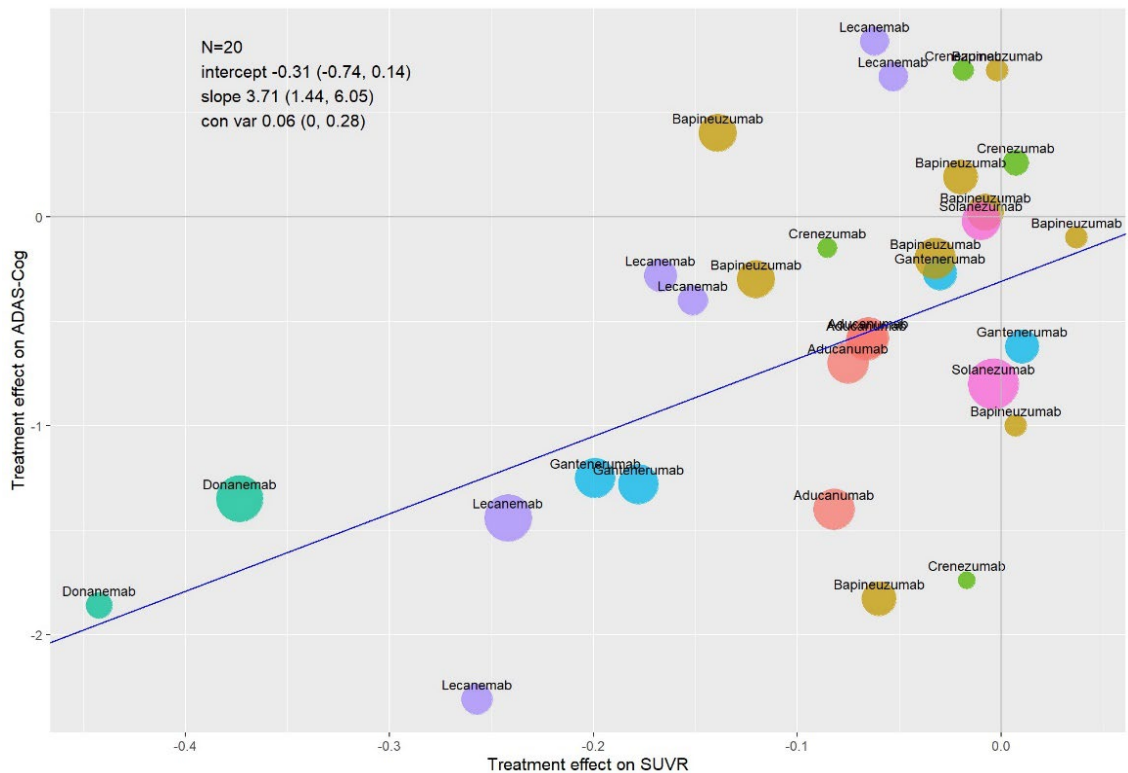

Figure S2. Bubble plot of the surrogate relationship between treatment effects on Aβ standardised uptake value ratio (SUVR) and Alzheimer's Disease Assessment Scale--Cognitive Subscale (ADAS-Cog) with treatment effects on Aβ reported at earlier time points. The mean (95% credible intervals) of intercept, slope and conditional variance were obtained from Daniels and Hughes model. The bubble size corresponds to the number of patients with measured cognitive functions.

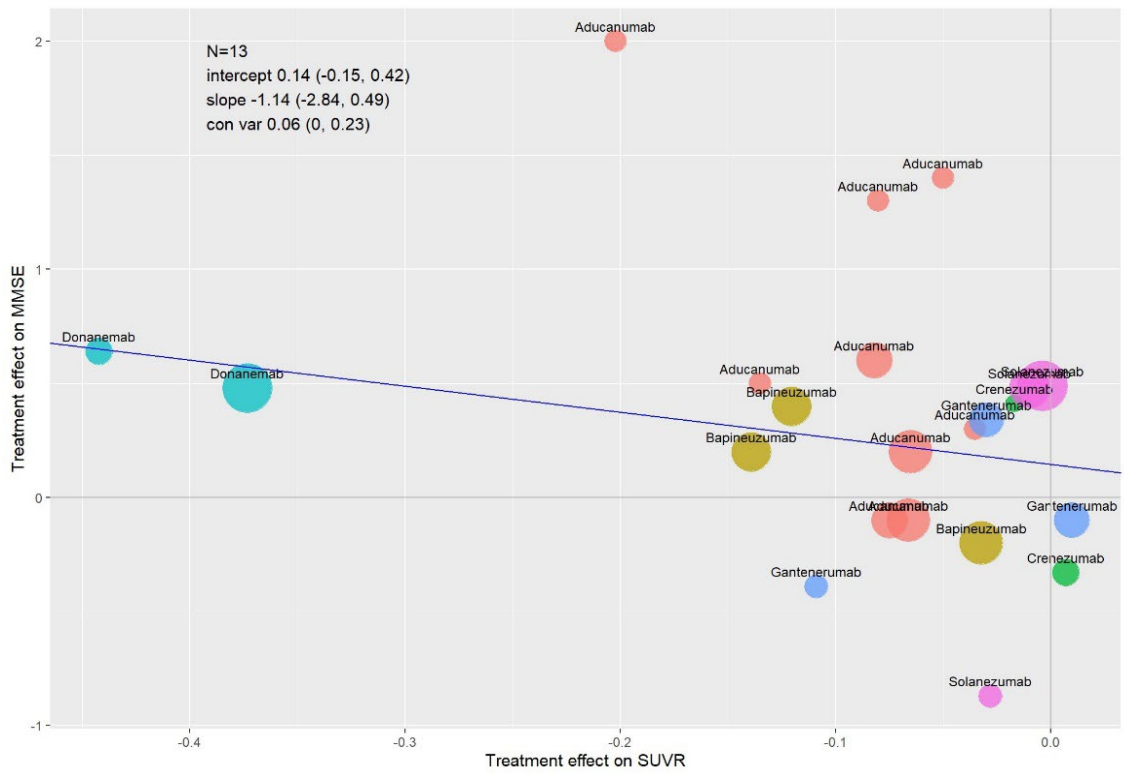

Figure S3. Bubble plot of the surrogate relationship between treatment effects on Aβ standardised uptake value ratio (SUVR) and Mini Mental State Examination (MMSE) with treatment effects on Aβ reported at earlier time points. The mean (95% credible intervals) of intercept, slope and conditional variance were obtained from Daniels and Hughes model. The bubble size corresponds to the number of patients with measured cognitive functions.

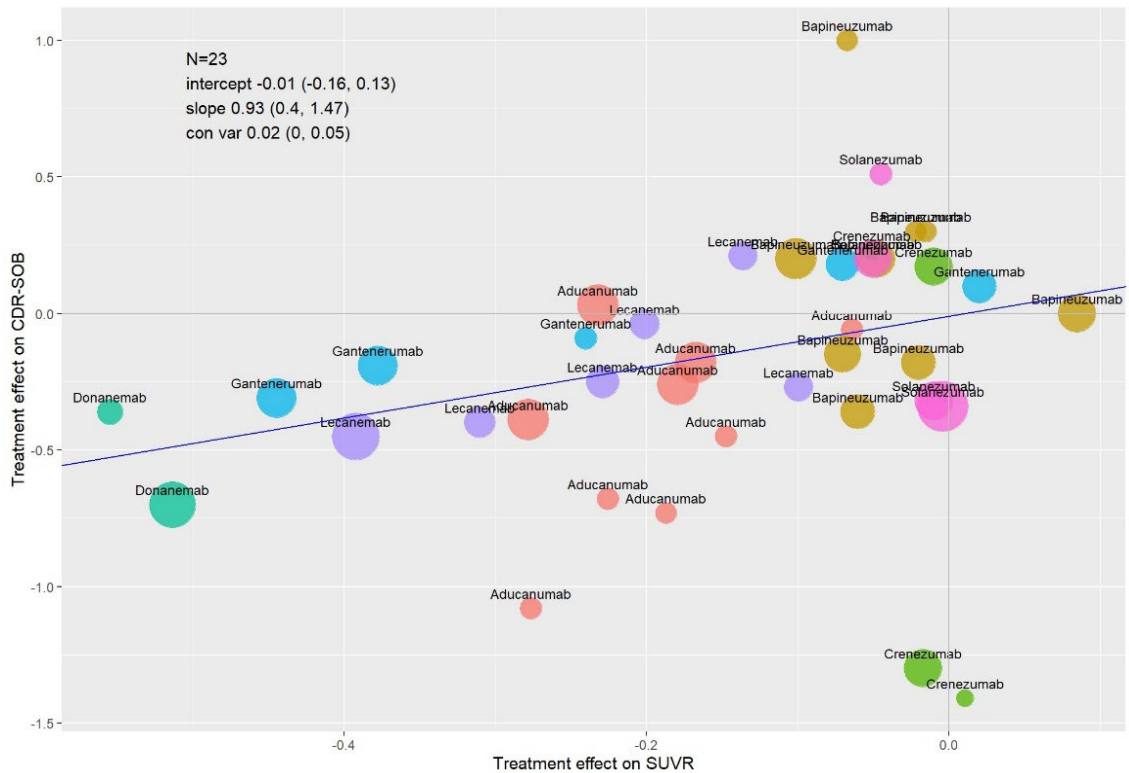

Figure S4. Bubble plot of the overall surrogate relationships with treatment effects on Aβ standardised uptake value ratio (SUVR) and Clinical Dementia Rating – Sum of Boxes (CDR-SOB) reported at the same time. The mean (95% credible intervals) of intercept, slope and conditional variance were obtained from Daniels and Hughes model. The bubble size corresponds to the number of patients with measured cognitive functions.

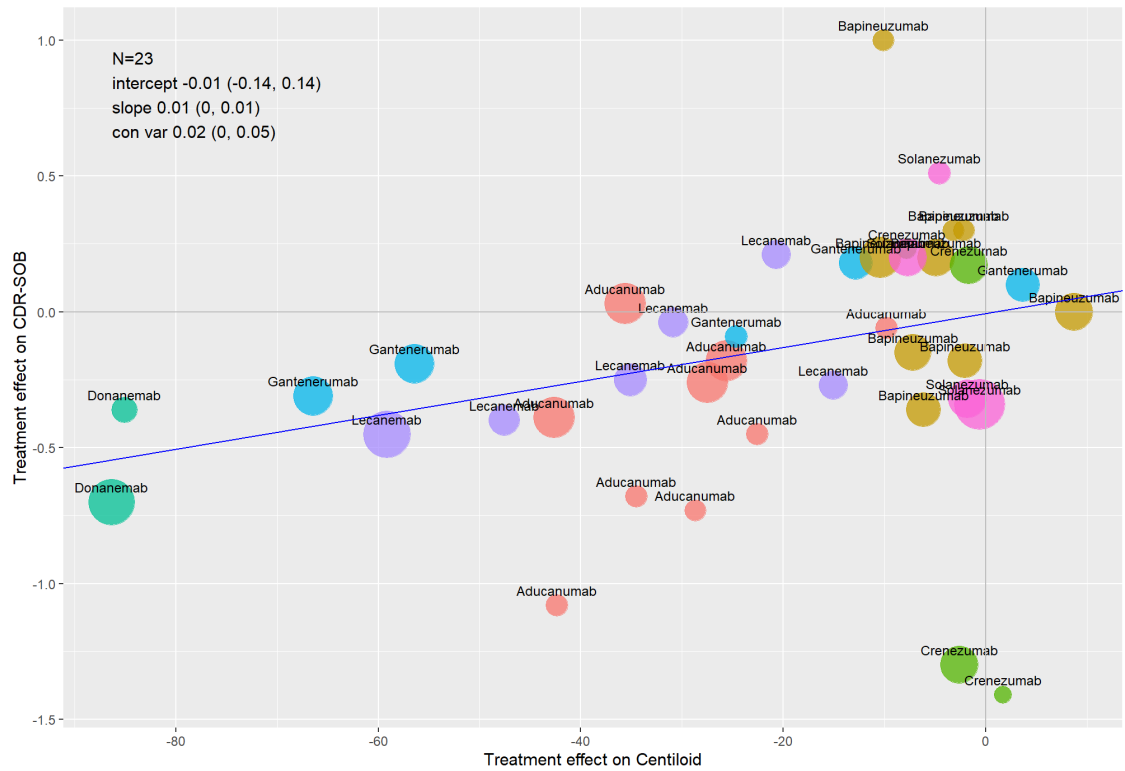

Figure S5. Bubble plot of the overall surrogate relationships with treatment effects on Aβ centiloids (with converted values) and Clinical Dementia Rating – Sum of Boxes (CDR-SOB) reported at the same time. The mean (95% credible intervals) of intercept, slope and conditional variance were obtained from Daniels and Hughes model. The bubble size corresponds to the number of patients with measured cognitive functions.

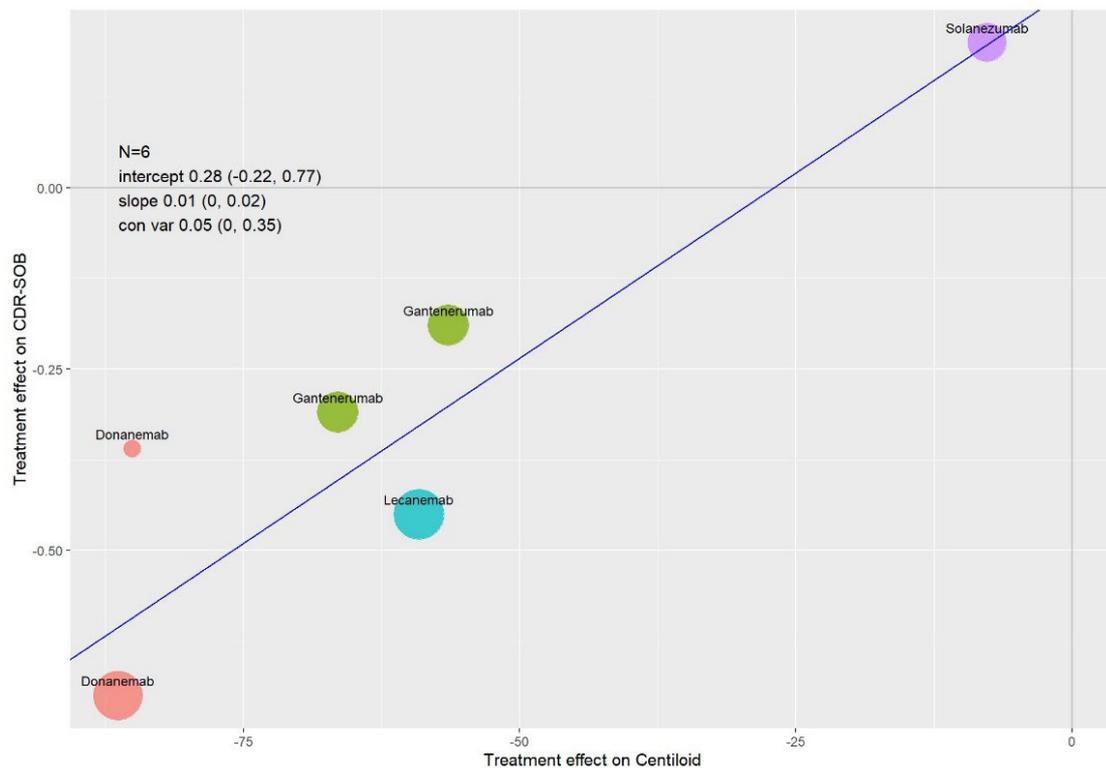

Figure S6. Bubble plot of the overall surrogate relationships with treatment effects on Aβ centiloids (without converted values) and Clinical Dementia Rating – Sum of Boxes (CDR-SOB) reported at the same time. The mean (95% credible intervals) of intercept, slope and conditional variance were obtained from Daniels and Hughes model. The bubble size corresponds to the number of patients with measured cognitive functions.

Appendix C3. Further analyses for evaluating surrogate relationship by treatment

The surrogate relationships for individual treatments were discussed in Section 3.2.1. The results of partial exchangeability model are presented along with the results of subgroup analysis and full exchangeability model in Figure S7.

The partial exchangeability hierarchical model allows for the flexible borrowing of information across treatments, resulting in the estimation of key parameters with less uncertainty. The surrogate relationship for lecanemab from the partial exchangeability hierarchical model was moderate, with the slope 1.71 (95% CrI: 0.08, 3.35) and conditional variance 0.03 (95% CrI: 0, 0.22). The surrogate relationship for all the other drugs was weak with zero included in the credible intervals of slope. For example, for aducanumab, the slope was 2.59 (95% CrI: -0.03, 6.92) and conditional variance was 0.03 (95% CrI: 0, 0.16).

The posterior mean of mixture weights from the partial exchangeability model had an average of 0.83, indicating a high level of borrowing of information. When comparing results from the full exchangeability model with subgroup analyses, the reduction in the width of CrI was 71% (51%-95%) for slope and 28% (7%-65%) for conditional variance. When comparing results from the partial exchangeability model with subgroup analyses, the reduction in the width of CrI was 54% (34%-94%) for slope and 24% (6%-63%) for conditional variance. The reduction is greater for the full exchangeability model as more information was borrowed in the full exchangeability model.

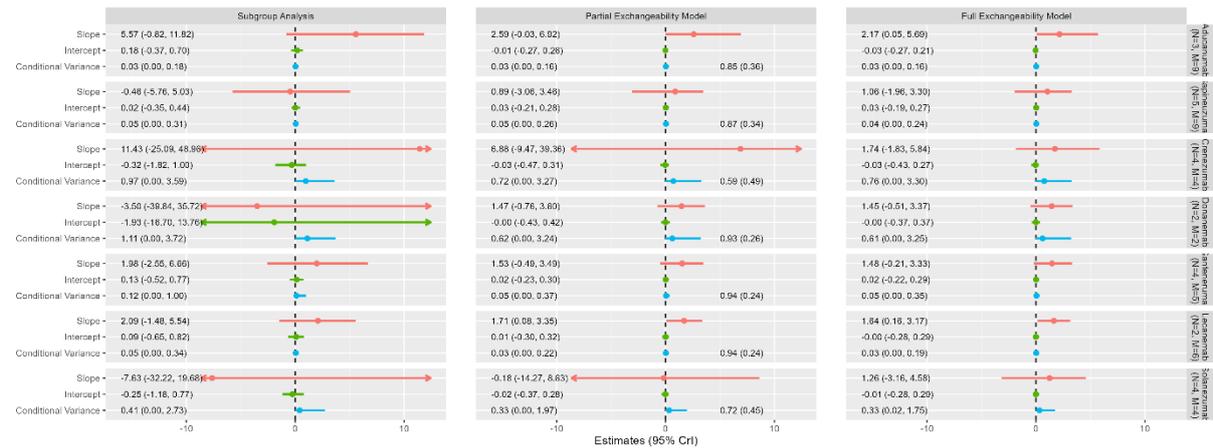

Figure S7. Forest plot of estimates of slope (red), intercept (green) and conditional variance (blue) for Aβ standardised uptake value ratio (SUVR) as a surrogate for Clinical Dementia Rating – Sum of Boxes (CDR-SOB). Each column represents a different model, and each row corresponds to a different treatment. N represents the number of studies included and M represent the total number of contrasts included.

Appendix C4. Cross validations and predictions for the analysis of surrogate relationships for individual treatments

Leave-one-out cross-validation was also performed to evaluate the predictive value of Aβ for CDR-SOB using the hierarchical models. The full exchangeability model demonstrated good coverage in terms of all predicted intervals including the observed effect estimates on CDR-SOB. The average absolute difference between the observed and predicted effect estimates was 0.32, and the average ratio of the width of intervals between the predicted and observed treatment effect was 2.49. A forest plot showing the observed effects and predicted effects of CDR-SOB for each study are presented in Figure S8.

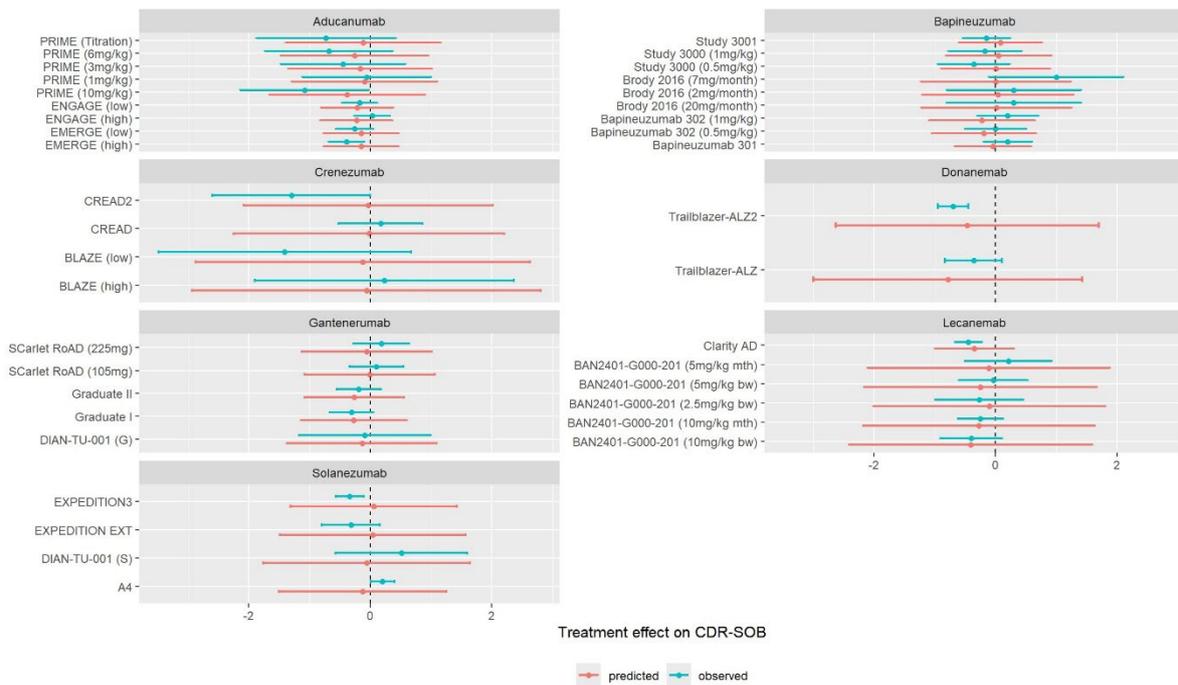

Figure S8. Forest plot illustrating the observed treatment on Clinical Dementia Rating – Sum of Boxes (CDR-SOB) and the corresponding predicted effect on CDR-SOB using full exchangeability model.